\documentclass[aps,pra,twocolumn,amsmath,amssymb]{revtex4}
\usepackage{graphicx}
\usepackage{dcolumn}
\usepackage{bm}
\usepackage{amsmath} 
 
\begin{document} 

\title{Bose-Fermi Mixtures in Optical Lattices}

\author{M. Centelles,
M. Guilleumas, M. Barranco, R. Mayol, and M. Pi} 

\affiliation{Departament d'Estructura i Constituents de la Mat\`eria,
Facultat de F\'{\i}sica, \\
Universitat de Barcelona, 
Diagonal {\sl 647}, {\sl 08028} Barcelona, Spain}


\begin{abstract}
Using mean field theory, we have studied Bose-Fermi mixtures in a
one-dimensional optical lattice in the case of an attractive
boson-fermion interaction. We consider that the fermions are in the
degenerate regime and that the laser intensities are such that quantum
coherence across the condensate is ensured. We discuss the effect of
the optical lattice on the critical rotational frequency for vortex
line creation in the Bose-Einstein condensate, as well as how it
affects the stability of the boson-fermion mixture. A reduction of the
critical frequency for nucleating a vortex is observed as the strength
of the applied laser is increased. The onset of instability of the
mixture occurs for a sizeably lower number of fermions in the presence
of a deep optical lattice.
\end{abstract}


\maketitle


\section{Introduction}
\label{Intro}

One of the most essential differences predicted between a superfluid
and a normal fluid is the eventual appearance of quantized vortices in
the former when set under rotation \cite{don91}. Not in vain, since
the demonstration of the occurrence of Bose-Einstein condensation
\cite{anderson95,bradley95,davis95}, the nucleation of vortices in
dilute ultra-cold atomic gases in rotating traps has been instrumental
to elucidate the superfluid properties of these systems and has
attracted much experimental and theoretical research (see e.g.\ Refs.\
\cite{dal96,fet01,theory1,theory2,theory3,theory4,experiment1, 
experiment2,experiment3} and references quoted therein). The
prominence of the study of vortices in atomic gases transcends the
frontiers of the field itself, because topological defects are
cornerstones in cross-disciplinary areas of physics. In turn, very
recently, the experimentation with Bose-Einstein condensates in
optical lattices is stimulating new avenues in the investigation of
quantum coherence and interference phenomena.

In the present work we study some properties of a slowly rotating
Bose-Einstein condensate (BEC) in a one-dimensional (1D) optical
lattice \cite{multivor}. We are motivated by the recent advances
achieved in the experimental manipulation of rotating condensates
\cite{ros02,bre04}, on the one hand, and of condensates loaded into
optical lattices \cite{had04}, on the other hand. Rotating ultracold
Bose gases in optical lattices are suitable systems to study quantum
phenomena. In particular, the excitations, stability and dynamics of a
vortex line in the superfluid regime have been recently addressed
\cite{vortex1OL}. Concurrently, the physics of degenerate Bose-Fermi
mixtures in optical lattices is attracting conspicuous attention.
Theoretical progress is being made in the basic understanding of the
system's quantum phase diagram and the superfluid to Mott-insulator
transition \cite{BFinOL}, and experimentalists have already been able
to prepare a degenerate mixture of Rb and K atoms in a tight optical
lattice \cite{mod03}.

The aim of this work is twofold. Firstly, we wish to address the
formation of vortices in a coherent array of Bose-Einstein condensates
in a 1D optical lattice. In doing this, we will determine the
dependence  of the thermodynamic critical angular velocity $\Omega_c$
for vortex formation \cite{dal96,fet01} on the laser intensity.
Secondly, we want to investigate the effect of trapped fermions on
$\Omega_c$. Attractive Bose-Fermi mixtures may experience collapse
and thus, we will also discuss the impact of the optical lattice on
the mechanical stability of the mixture.

The manuscript has been organized as follows. In Section \ref{Theory}
a short description of the mean field model for the present problem is
given. Section \ref{BEC} is devoted to the discussion of the numerical
results for a rotating boson condensate confined by magnetic and
optical traps. In Section \ref{Mixture} we address the properties of a
$^{87}$Rb--$^{40}$K mixture loaded in an optical lattice. Our
concluding remarks are laid in Section \ref{Conclusion}. Finally, a
derivation of the virial theorem for the mixture in the 1D optical
lattice is given in the Appendix.

\section{Theory}
\label{Theory}

We consider a zero temperature mixture made of a $^{87}$Rb Bose
condensate (B) and a degenerate $^{40}$K Fermi gas (F). They are
confined by the axially-symmetric external potentials of a harmonic
magnetic trap and of a stationary periodic optical lattice modulated
along the $z$-axis. The lattice is produced by a far detuned laser,
which hinders the possibility of spontaneuous scattering and yields
practically equal Rb and K optical potentials. The resulting potential
for each kind of atom $q=B,F$ is
\begin{equation}
V_q = \frac{1}{2}\,m_q (\omega_{q \perp}^2 r^2 + \omega_{q z}^2 z^2)+
        \frac{V_0}{2} \cos (2\pi z / d) \,,
\label{vextBF}
\end{equation}
where $m_q$ is the atomic mass and $r= \sqrt{x^2+y^2}$ is the radial
variable of cylindrical coordinates.

The radial and axial frequencies of the harmonic trap are taken from a
recent experiment \cite{had04}: $\omega_{B \perp}= 2 \pi \nu_{B
\perp}$ with $\nu_{B \perp}= 74$ Hz, and $\omega_{B z}= 2 \pi \nu_{B
z}$ with $\nu_{B z}= 11$ Hz for $^{87}$Rb, while those for $^{40}$K
are a factor $(m_B/m_F)^{1/2} \simeq 1.47$ larger. The optical
potential is determined by its period $d=2.7 \mu$m \cite{had04} and
depth $V_0=s E_R$, where $E_R=\hbar^2 \pi^2/2md^2=h \times 80$ Hz is
the recoil energy and $s$ is a dimensionless parameter that provides
the intensity of the laser beam.

For $3 \times 10^5$ atoms of $^{87}$Rb confined in the harmonic trap,
the condensate is cigar-shaped with a Thomas-Fermi length $L_{\rm
TF}=84 \,\mu$m, and a radius $R_{\rm TF}= 6 \, \mu$m. The superimposed
optical potential splits the BEC over $L_{\rm TF}/d \sim 30$ wells. We
shall be concerned with a range of laser intensities that ensures full
coherence of the condensate across the whole system \cite{vortex1OL}.
The criterion for the Mott transition \cite{zwe03} for the considered
parameters leads to an estimate of a maximum lattice depth for the
superfluid regime of $V_0 \sim 150 \, E_R \sim h \times 12$ kHz
\cite{had04}. Provided that the strength $V_0$ does not surpass
largely this value, quantum tunneling between consecutive wells is
still sufficient to retain coherence and one can apply the
Gross-Pitaevskii theory for the order parameter to study the
properties of the condensate.
                                                                  
Within the mean-field approach, neglecting $p$-wave interactions, the
energy density functional that describes the boson-fermion mixture at
zero temperature with a quantized vortex in the condensate along the
$z$-axis has the form \cite{ro02,jez04}
\begin{eqnarray}
{\cal E}(\textbf{r}) &=& \frac{ \hbar^2 }{2 m_B} \left[
        (\mbox{\boldmath$\nabla$} n_B^{1/2} )^2+
        \frac{\kappa^2}{r^2} n_B \right] +
        V_B \,n_B 
       + \frac{1}{2} \, g_{BB}\, n_B^2
       \nonumber\\[1mm]
       & & \null
+ \frac{\hbar^2}{2 m_F} 
 \left[ \frac{3}{5} (6 \pi^2)^{2/3} \, n_F^{5/3}
 + \frac{1}{36} \frac{(\mbox{\boldmath$\nabla$} n_F)^2}{n_F} \right]
       \nonumber\\[1mm]
       & & \null 
       + V_F \, n_F + g_{BF} \,n_F\, n_B \,,
       \label{ed}
\end{eqnarray}
where $\kappa$ is the circulation quantum number, $n_B= |\Psi|^2$ is
the condensate density, $n_F$ is the fermion density, and the fermion
kinetic energy density has been written in the
Thomas-Fermi-Weizs\"acker approximation. The wave function of the
boson condensate with a vortex line is $\Psi= \sqrt{n_B(r,z)} \, e^{i
\kappa \phi}$, where $\phi$ refers to the azimuthal angle in the
xy-plane. In this work we consider the cases $\kappa=0$ (non-rotating
condensate) and $\kappa=1$ (condensate with a singly quantized
vortex). As is known, vortices with more than one quantum of
circulation will not exist in equilibrium \cite{theory3}. The
boson-boson and boson-fermion coupling constants $g_{BB}$ and $g_{BF}$
that enter Eq.~(\ref{ed}) are written in terms of the $s$-wave
scattering lengths $a_B$ and $a_{BF}$ as
$g_{BB}=4\pi\,a_{B}\hbar^2/m_B$ and
$g_{BF}=4\pi\,a_{BF}\hbar^2/m_{BF}$, respectively, with $m_{BF} \equiv
2 m_B m_F / (m_B + m_F)$.

The variation of ${\cal E}$ with respect to $n_B$ and $n_F$, under the
constraint of given number of bosons $N_B$ and fermions $N_F$, yields
two coupled Euler-Lagrange equations. Namely, a Gross-Pitaevskii
equation for bosons with a term describing the boson-fermion
interaction: 
\begin{eqnarray}
 \left\{ \frac{\hbar^2}{2 m_B}
 \left[ - \Delta + \frac{\kappa^2}{r^2} \right]
 + V_B + g_{BB} n_B + g_{BF} n_F \right\} n_B^{1/2}
& & 
\nonumber \\[1mm]
\, = \, \mu_B \, n_B^{1/2} ,
\qquad
\label{gp}
\end{eqnarray}
and a Thomas-Fermi-Weizs\"acker equation for fermions
\cite{ro02,jez04} which is increasingly valid the larger $N_F$ is:
\begin{eqnarray}
\bigg\{ \frac{\hbar^2}{2 m_F}
        \left[(6 \pi^2)^{2/3} \, n_F^{2/3} + \frac{1}{36}
        \frac{(\nabla n_F)^2}{n_F^2} - \frac{1}{18} \Delta \right]
& & 
\nonumber \\[1mm]
\mbox{}
+ V_F +  g_{BF}\, n_B \bigg\} \, n_F \, = \, \mu_F \, n_F,
\qquad
\label{tf}
\end{eqnarray}
where $ \mu_B$ and $ \mu_F$ are the chemical potentials of the boson
and fermion atomic species, respectively. The ground state of the
boson-fermion mixture is found with $\kappa=0$, while the vortical
state is computed with $\kappa=1$. Starting from randomly sampled
densities, we solve Eqs.~(\ref{gp}) and (\ref{tf}) by means of the
imaginary time method \cite{bar03}. We use the adapted version of the
virial theorem \cite{jez04} to check the numerical convergence of the
solution (see the Appendix).

\section{Boson Condensate Loaded into an Optical Lattice}
\label{BEC}

\begin{figure}
\includegraphics[width=0.78\columnwidth,
angle=270, clip=true]{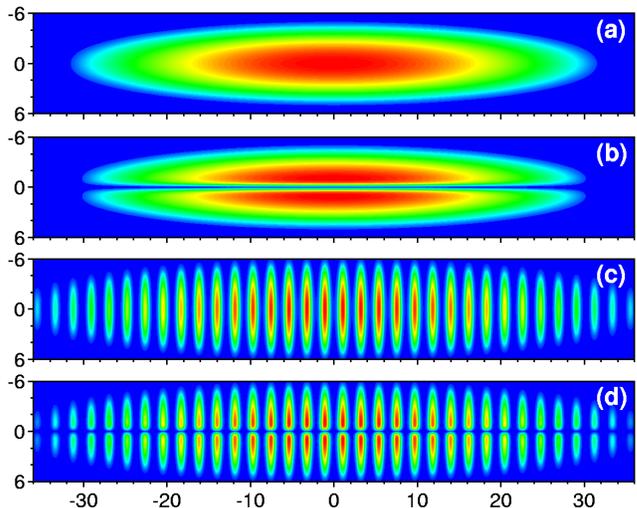}
\caption[]{(Color online) 
Contour plots of the $^{87}$Rb condensate density in the $xz$-plane.
(a) Condensate in the pure harmonic trap.
(b) Same as (a) with a quantized vortex along the $z$-axis.
(c) Same as (a) under an optical lattice with $V_0/h =2.5 \,$kHz. 
(d) Same as (c) with a quantized vortex along the $z$-axis.
Distances are in units of the oscillator length 
$a_{B \perp}= \sqrt{\hbar/(m_B \omega_{B \perp})} = 1.25 \mu$m.
In each panel the color scale ranges from deep blue at lowest
densities to red at highest densities [these are 229 (a), 212 (b), 394
(c), and $376 \,\, a_{B \perp}^{-3}$ (d)].}
\label{fig1}
\end{figure}

We first consider a pure boson condensate, with $N_B=3 \times 10^5$
atoms of $^{87}$Rb as in the experiment of Ref.~\cite{had04}. For the
boson-boson scattering length we use the value $a_B=98.98 \, a_0$ ($1
\, a_0 = 0.529 \,\mbox{\AA}$) \cite{kem02}. In Fig.~\ref{fig1} we show
contour plots of the condensate density in the $xz$-plane,
$n_B(x,0,z)=|\Psi(x,0,z)|^2$, for different situations. In (a) the
condensate is confined only by the harmonic trap. In (b) the
cigar-shaped condensate of (a) hosts a quantized vortex line along the
$z$-axis. In (c) a 1D optical lattice with $V_0/h = 2.5 \,$kHz splits
 the cigar-shaped condensate (a) into an array of multiple disk-like
coherent condensates. 

As shown in Ref.\ \cite{kra02}, certain aspects
of the macroscopic properties and low-energy dynamics of a
magnetically trapped BEC in a tight optical lattice can be understood
in terms of a renormalized interaction coupling constant $g^*_{BB} >
g_{BB}$ and of an effective mass $m^*_B > m_B$ along the direction of
the periodic optical potential. The increase of the radial size of the
sample that one observes in panel (c) with respect to panel (a) is due
to the increased repulsive effect of the boson-boson interaction when
the system feels the optical lattice. It originates from the fact that
the optical confinement produces a local compression of the gas inside
each well \cite{zwe03,kra02}. The axial size of the condensate also
increases appreciably due to the combined effect of the repulsive
interactions and the redistribution of atoms inside the potential
wells. 

In panel (d) of Fig.~\ref{fig1} the optical lattice is superimposed to
the condensate with the vortex state (b). In a bulk superfluid, the
size of the vortex core is of the order of the healing length $\xi=[8
\pi n_B a_B]^{-1/2}$, where $n_B$ is the bulk density. This expression
holds for inhomogeneous superfluids taking for $n_B$ the local boson
density in the absence of vortices \cite{fet01}. Thus, the vortex core
size is larger for the outer sites of the split condensate, where the
density is smaller. 

\begin{figure}
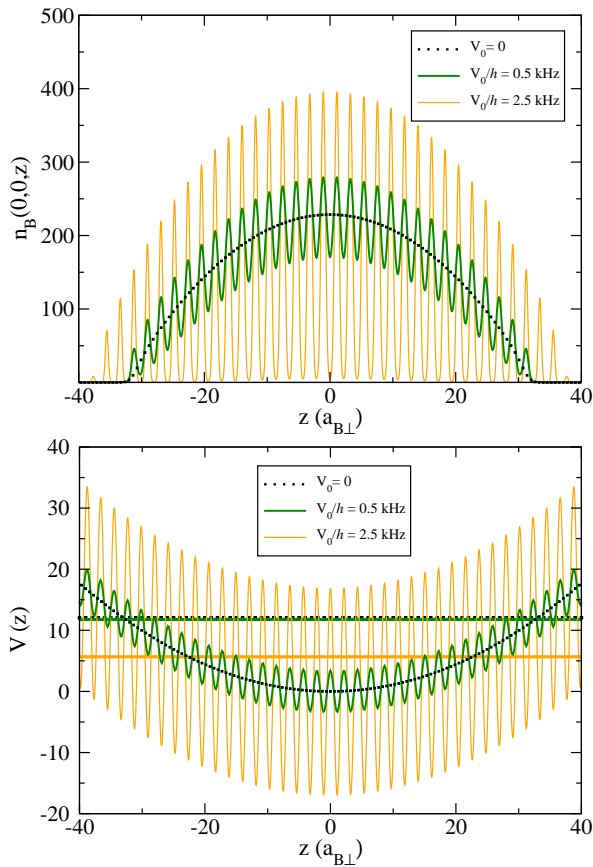

\includegraphics[width=0.90\columnwidth,
angle=0, clip=true]{opticalBF_lphys_f02a.eps}
\includegraphics[width=0.90\columnwidth,
angle=0, clip=true]{opticalBF_lphys_f02b.eps}
\caption[]{(Color online) 
Top: Density profile of the condensate along the $z$-axis
$n_B(0,0,z)=|\Psi(0,0,z)|^2$, in units of $a_{B \perp}^{-3}$, for
the indicated values of the lattice depth $V_0$.
Bottom: The $z$-dependence of the combined potential of the harmonic
trap and the optical lattice, Eq.~(\ref{vextBF}), in units of
$\hbar \omega_{B \perp}$.
The horizontal lines depict the location the chemical
potential of the condensate boson cloud trapped in that potential.
}
\label{fig2}
\end{figure}

In the upper part of Fig.~\ref{fig2} we show the evolution of the
density profile of the condensate along the $z$-axis,
$n_B(0,0,z)=|\Psi(0,0,z)|^2$, with the laser intensity. In the lower
part of Fig.~\ref{fig2} we depict the total trapping potential felt by
the atomic condensate. In the presence of a shallow 1D optical
lattice, the condensate profile starts developing small oscillations
around the profile at $V_0=0$ which follow the periodic optical
potential, but its spatial extension is nearly not affected. When the
laser intensity increases the system nearly splits into separate
condensates. The effect of the local compression inside each potential
well of the lattice, as well as the increase of the axial size of the
system, can be clearly seen in Fig.~\ref{fig2}.

\begin{figure}
\includegraphics[width=0.90\columnwidth,
angle=0, clip=true]{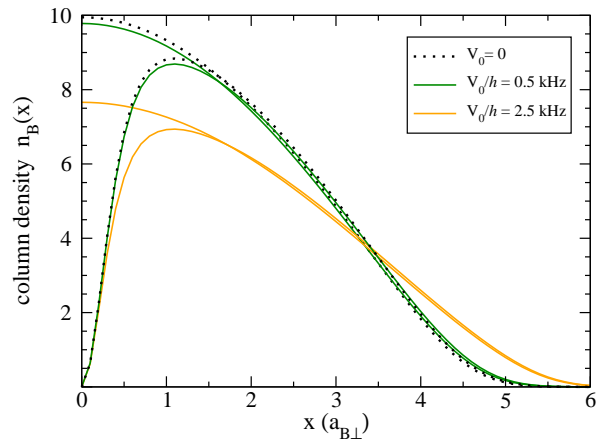}
\caption[]{(Color online) 
Column density $n_B(x)$, in units of $10^3 \times a_{B \perp}^{-2}$,
for different laser intensities. The column density of the condensate
when it hosts a vortex line is also plotted.
}
\label{fig3}
\end{figure}

In Fig.~\ref{fig3} we plot the column density $n_B(x)=\int dz \,
n_B(x,0,z)$ for the condensate in the harmonic trap alone and when the
optical lattice is added. A deep optical lattice results in a visible
reduction of the column density at the interior of the condensate and
in an extended surface. Figure~\ref{fig3} also displays $n_B(x)$ for
the condensate hosting a vortex line. In this case the quantized
circulation around the vortex line pushes the atoms away from it.

\begin{figure}
\includegraphics[width=0.90\columnwidth,
angle=0, clip=true]{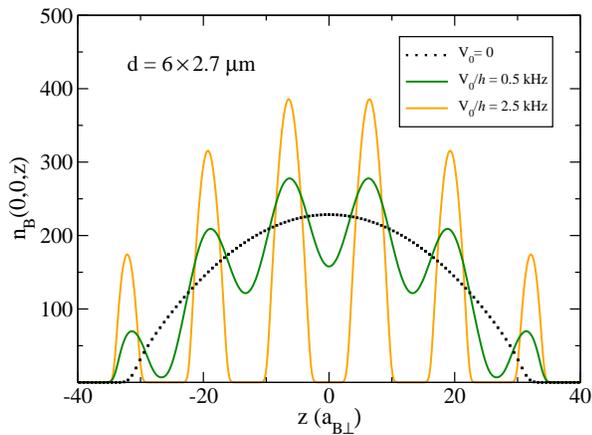}
\caption[]{(Color online) 
Density profile of the condensate along the $z$-axis
$n_B(0,0,z)=|\Psi(0,0,z)|^2$, in units of $a_{B \perp}^{-3}$, for
different laser intensities when the period of the optical lattice
potential is increased to $d= 6 \times 2.7 \mu$m.
}
\label{fig4}
\end{figure}

The effect of the modification of the value of the lattice constant
$d$ is illustrated in Fig.~\ref{fig4}. In this figure we plot the
computed density profile $n_B(0,0,z)$ of the boson atoms in the
condensate for the same parameters as in Fig.~\ref{fig2} but for a
larger $d$ ($d= 6 \times 2.7 \mu$m).


\begin{figure}
\includegraphics[width=0.90\columnwidth,
angle=0, clip=true]{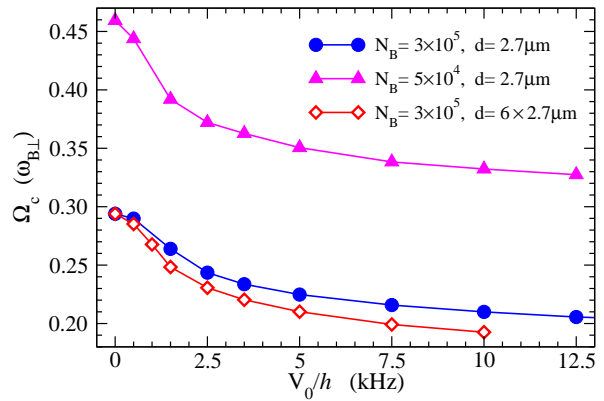}
\caption[]{(Color online) 
Critical angular velocity in units of $\omega_{B \perp}$ for
nucleation of a singly quantized vortex line as a function of the
laser intensity.
}
\label{fig5}
\end{figure}

The thermodynamic critical angular velocity $\Omega_c$ for nucleating
a singly quantized vortex is obtained by subtracting from the vortex
state energy $E_1$ in the rotating frame the ground-state energy
$E_0$, i.e., $\Omega_c=(E_1/N_B -E_0/N_B)/\hbar$ \cite{dal96,fet01},
and it provides a lower bound to the critical angular velocity
\cite{fet01}. In Fig.~\ref{fig5} we plot $\Omega_c$ as a function of
the laser intensity for $N_B= 3 \times 10^5$ and for $N_B= 5 \times
10^4$. For a fixed number of condensate atoms, $\Omega_c$ decreases
when $V_0$ increases, in agreement with the fact that the radial size
of the system becomes larger in the presence of the optical lattice
and the associated reduction of atoms along the symmetry axis (see the
decrease of the column density $n_B$ at the origin in Fig.~\ref{fig3}
with increasing $V_0$). In turn, the increase of the inertia of the
condensate along the direction of the laser beam due to the larger
effective mass $m^*_B$ \cite{kra02} also contributes to the reduction
of the critical angular velocity.

Assuming that the relative effect remains of the same order in
experiment, Fig.~\ref{fig5} predicts a sizeable $24\%$ reduction
of $\Omega_c$ already with a laser strength $V_0/h=5$ kHz for both of
the $N_B$ values. The effect is enhanced the shallower the magnetic
trap is. A less elongated magnetic trap would favor the appearence of
a vortex at a slower rotation. For example, the values of 
$\Omega_c(V_0=0)$ shown in Fig.~\ref{fig5} would be decreased by a 0.6
factor in a spherical trap with $\nu_{B z}= \nu_{B \perp}= 74$ Hz.
The relative reduction of $\Omega_c$ caused by $V_0/h=5$ kHz
with respect to $V_0=0$ would still be of 20--23\%.
As the thermodynamic $\Omega_c$ underestimates the actual critical
frequency for vortex creation, dynamical calculations would be needed
for a precise quantitative estimate of these changes.

Figure~\ref{fig5} suggests an experiment to test the dependence of the
critical frequency for vortex nucleation on the intensity of the
optical lattice, since it shows that, for a rotating condensate with a
given $N_B$, a vortex line should nucleate at angular frequencies
much lower than $\Omega_c(V_0=0)$ if a co-rotating deep 1D optical
lattice is superimposed, whereas otherwise it would not. The
experimental observation of this effect would provide a clear
signature of the correctness of the physics predicted by the
Gross-Pitaevskii theory for rotating condensates in optical lattices.

\section{Attractive Boson-Fermion Mixture Loaded into an Optical
Lattice}
\label{Mixture}

With the purpose of assessing the effects of a trapped fermion cloud
on the rotating condensate in an optical lattice, we consider next a
$^{87}$Rb--$^{40}$K mixture. It is characterized by a large attractive
Bose-Fermi interaction that assists the sympathetic cooling of the
fermionic species down to the degenerate regime. There are various
values of the interspecies scattering length $a_{BF}$ published,
depending on the type of experiment and techniques used
\cite{fer02,mod02,modcol03,gol04}. A recent measurement of $a_{BF}$ at
JILA \cite{gol04} has established a value $a_{BF}=-250 \pm 30 \, a_0$,
which we adopt here because of its reduced experimental uncertainties.
Vortex states in $^{87}$Rb--$^{40}$K mixtures confined by a harmonic
potential have been recently addressed \cite{jez04} (although
$a_{BF}=-395 \, a_0$ \cite{modcol03} was employed). When a 1D optical
lattice is switched on, the atomic species experience the trapping
potential (\ref{vextBF}).

We have carried out calculations for a mixture with $N_B=3 \times
10^5$ and $N_F=1.5 \times 10^5$ atoms, and the same parameters as in
Fig.~\ref{fig1}. We assume that the Fermi component is in the
normal---nonsuperfluid---but quantum degenerate phase, and consider
that it is in a stationary state. This situation could be achieved
experimentally by waiting long enough after the generation of the
vortex in the condensate to let the drag force between bosons and
fermions to dissipate.

\begin{figure}
\includegraphics[width=0.78\columnwidth,
angle=270, clip=true]{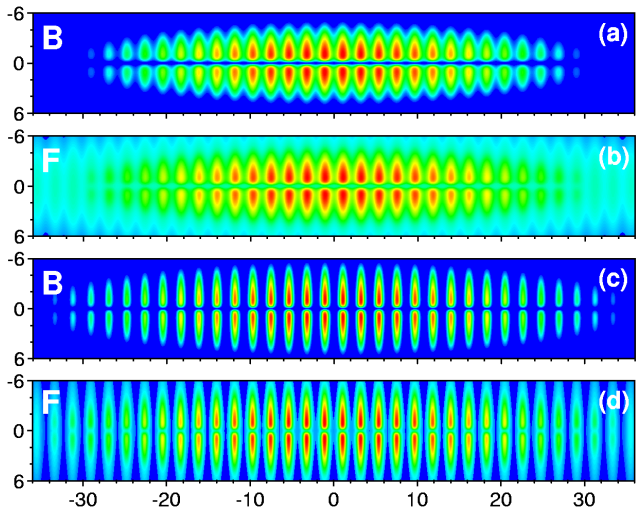}
\caption[]{(Color online) 
Contour plots of the condensate (B) and fermionic (F) densities in the
$xz$-plane for the $^{87}$Rb--$^{40}$K mixture, in the combined
harmonic and optical lattice trap. The condensate hosts a quantized
vortex. In (a) and (b) $V_0/h = 0.5 \,$kHz, whereas in (c) and (d)
$V_0/h = 2.5 \,$kHz. Distances are in units of $a_{B \perp}$.
In each panel the color scale ranges from deep blue at lowest
densities to red at highest densities [these are 312 (a), 25 (b), 457
(c), and $37 \,\, a_{B \perp}^{-3}$ (d)].}
\label{fig6}
\end{figure}

We present in Fig.~\ref{fig6} contour plots of the
boson $n_B(x,0,z)$ and fermion $n_F(x,0,z)$ densities when the
condensate hosts a vortex line, for two laser strengths: $V_0/h = 0.5$
(a,b) and $2.5$ kHz (c,d). The condensate and the fermionic cloud are
modulated by the regular pattern of the optical lattice. The large
mutual attraction between fermions and bosons makes the effective
interaction between bosons less repulsive than in the pure BEC.
Furthermore, the boson atoms induce an effective attraction between
fermions counteracting the Fermi pressure. As a result, the density of
both species increases in the overlapping region, as if they were more
strongly confined by the external potentials, and the condensate
becomes more compact in space (compare Figs.~\ref{fig6}c and
\ref{fig1}d). Still, the effect of the Pauli exclusion principle is
notorious in Fig.~\ref{fig6} where the $^{40}$K cloud is seen to
extend to larger distances than the $^{87}$Rb atoms, both axially and
radially. 

The presence of a vortex along the $z$-axis in the condensate
component is apparent in panels (a) and (c) of Fig.~\ref{fig6} as the
boson density vanishes on the vortex line. It is interesting to note
that the large attractive $a_{BF}$ leads to a visible depletion of the
fermionic  density on the $z$-axis as well, reminiscent of the bosonic
vortex  core. The effect should be directly observable experimentally.
It is more evident in (d) where bosons and fermions are more tightly
confined by the optical lattice.

In the present mixture we find that the presence of the quantum
degenerate fermionic cloud does not change much the value of
$\Omega_c$ compared to the pure BEC, nor its pattern against the
optical lattice strength. Indeed, for all the $V_0$ values considered
here the critical frequency for vortex appearence is raised by
$\sim 10\%$, due to the enhancement of the condensate density at the
core caused by the fermion atoms. The impact of the fermion cloud on
$\Omega_c$ is magnified in a deeper magnetic trap. For instance, if
the same mixture was set in a trap with $\nu_{B z}= 95$ Hz and $\nu_{B
\perp}= 640$ Hz, that preserves the aspect ratio $\nu_{B z}/\nu_{B
\perp}=0.15$ used in Fig.~\ref{fig6}, the critical frequency
$\Omega_c(V_0=0)$ would increase by $40\%$ with respect to the $N_F=0$
case. In this squeezing magnetic trap, however, the effect of $V_0$ on
$\Omega_c$ is negligible ($<2\%$).

A peculiar feature of the $^{87}$Rb--$^{40}$K system, that stems from
the interplay of the moderately repulsive $a_B$ and the strongly
attractive $a_{BF}$, is the existence of a mechanical stability limit
beyond which the system cannot sustain more fermions. At high
densities the strong interspecies attraction may overcome the Pauli
pressure and drive the ultracold gas mixture into collapse. The
phenomenon has been observed experimentally \cite{mod02}. This
motivates us to address next the stability of the $^{87}$Rb--$^{40}$K
system subject to an optical lattice. In fact, collapse is currently
the focus of experimental attention because of the implications the
instability has for constraining $a_{BF}$ \cite{mod02,modcol03,gol04},
and also because there are prospects that a radially-squeezed
$^{87}$Rb--$^{40}$K mixture at a density close to collapse must be
able to form stable bright soliton trains \cite{bon04}.

If we keep $N_B$ fixed at $3 \times 10^5$ atoms as in the previous
discussions, with $a_{BF}=-250 \, a_0$ we find that the boson-fermion
mixture confined in the magnetic trap of Fig.~\ref{fig6} would be
stable up to virtually arbitrary fermion numbers, consistently with
the findings of Ref.~\cite{ro02}. Increasing $\omega_{B \perp}$
prompts the occurrence of collapse. In a trap with $\nu_{B z}= 95$ Hz
and $\nu_{B \perp}= 640$ Hz, the $3 \times 10^5$ rubidium atoms are
able to retain up to a maximum of $N_F^{\rm max}= 2.6\times 10^5$
potassium atoms, when the system collapses. We recall that the effect
on $\Omega_c$ of $N_F=1.5\times 10^5$ and $V_0$ for this squeezing
trap has been discussed before.

Now, with $\nu_{B z}= 95$ Hz and $\nu_{B \perp}= 640$ Hz, a decrease
of $N_F^{\rm max}$ can be seen when a 1D optical lattice is added. It
is related to the increase of the effective local compression of the
atoms caused by the optical trap. Indeed, superimposing a 1D optical
lattice with $d=2.7 \mu$m and $V_0/h= 5$ kHz to this strongly
confining harmonic trap, the $^{87}$Rb--$^{40}$K mixture with
$N_B=3\times 10^5$ can sustain up to $N_F^{\rm max}= 7.5\times 10^4$
fermions only, i.e., less than 30\% of the $V_0=0$ value. The effect
appears to be accessible to verification under present experimental
conditions. It opens the possibility to study the collapse of a
mixture for trapped fermion and boson numbers considerably smaller
than in a pure harmonic trap.

\section{Conclusion}
\label{Conclusion}

Summarizing, we have determined the thermodynamic critical frequency
for nucleating a quantized vortex in an array of coherent condensates
within the Gross-Pitaevskii theory. Next we have studied a Bose-Fermi
mixture trapped in a 1D optical lattice to examine the effect on the
value of $\Omega_c$ of a fermionic cloud in the quantum degenerate
phase. $^{87}$Rb--$^{40}$K mixtures may collapse and, in this regard,
it has been shown that a 1D optical lattice can be an efficient degree
of freedom to tune the onset of instability. Our analysis of the
relative variation of $\Omega_c$ calculated by thermodynamic arguments
may provide a useful information for future experiments.

\section*{ACKNOWLEDGMENTS}

We thank Dr.\ K. Bongs for valuable discussions. This work has been
performed under Grants No.\ BFM2002-01868 from DGI (Spain) and FEDER,
and No.\ 2001SGR-00064 from Generalitat de Catalunya. M.G. thanks the
``Ram\'on y Cajal'' Program (Spain) for financial support.

\section*{APPENDIX}

In this Appendix we make use of the principle of scale
invariance to obtain the virial theorem for the energy functional of
the boson-fermion mixture in an optical lattice. The virial theorem
results from homogeneity properties of the kinetic and
potential components of the energy of the many-body system with
respect to a scaling transformation that preserves the normalization
\cite{parr89}. 

A normalized scaled version of the particle density, for either bosons
or fermions, is
\begin{equation}
n_{\lambda}(\textbf{r}) =
\lambda^3 \, n(\lambda \textbf{r}) \,,
\label{a1}
\end{equation}
where $\lambda$ is an arbitrary scaling parameter.
On account of (\ref{a1}), a straightforward manipulation shows
that the various contributions to the energy from the functional
(\ref{ed}) scale following the rules
\begin{equation}
T_{\lambda} =  \lambda^2 \, T
\end{equation}
for the kinetic energy (including the boson vortex term if present), 
\begin{equation}
U_{H \lambda} =  \frac{1}{\lambda^2} \, U_H
\end{equation}
for the harmonic potential terms, and
\begin{equation}
U_{g \lambda} =  \lambda^3 \, U_g 
 \quad \mbox{and} \quad
U_{g_{BF} \lambda} = \lambda^3 \, U_{g_{BF}}
\end{equation}
for the intra- and inter-species interaction terms. The contribution
of the optical lattice transforms as
\begin{eqnarray}
U_{\mathrm{opt} , \lambda} & = & \frac{V_0}{2}
\int d (\lambda\textbf{r}) 
\, \cos \bigg[ \frac{2\pi (\lambda z)}{\lambda d} \bigg]
\, n(\lambda\textbf{r}) 
\nonumber
\\[1mm]
& = & U_\mathrm{opt}(\lambda d) \,.
\end{eqnarray}

Hence, the total energy of the system scales as
\begin{eqnarray}
 E_{\lambda} & = & \lambda^2 (T_B + T_F)
 +\frac{1}{\lambda^2} (U_{H_B} + U_{H_F}) 
\nonumber
\\[1mm]
& & \mbox{}
 +  \lambda^3 ( U_{g_{BB}} + U_{g_{BF}}) 
\nonumber
\\[1mm]
& & \mbox{}
+ U_{\mathrm{opt}_B}(\lambda d) 
+ U_{\mathrm{opt}_F}(\lambda d) \,.
\label{a3}
\end{eqnarray}
And, finally, the stationarity condition of the energy against
dilation leads to the virial theorem for the boson-fermion mixture:
\begin{eqnarray}
0 & = & \left.\frac{d E_{\lambda}}{d \lambda}\right|_{\lambda=1} 
\nonumber
\\[1mm]
& =  & 
2 (T_B + T_F) - 2 (U_{H_B} + U_{H_F} )
\nonumber
\\[1mm]
& & \mbox{}
+ 3 ( U_{g_{BB}} + U_{g_{BF}}) 
\nonumber
\\[1mm]
& & \mbox{}
 + \frac{1}{d} \bigg[ \frac{\partial U_{\mathrm{opt}_B}}{\partial d}
 +  \frac{\partial U_{\mathrm{opt}_F}}{\partial d} \bigg] \,,
\label{a4}
\end{eqnarray}
where
\begin{eqnarray}
\lefteqn{  \frac{1}{d} \bigg[ \frac{\partial
U_{\mathrm{opt}_B}}{\partial d}
 +  \frac{\partial U_{\mathrm{opt}_F}}{\partial d} \bigg]
}
& &
\nonumber
\\[1mm]
& = &
 \frac{V_0}{2} \int d \textbf{r} 
 \sin \bigg[ \frac{2\pi z}{d} \bigg] \frac{2\pi z}{d}
 \, [n_B(\textbf{r}) + n_F(\textbf{r}) ] .
\label{a5}
\end{eqnarray}

Equation~(\ref{a4}) is a useful tool to check the accuracy of the
numerical solution of the coupled Gross-Pitaevskii and
Thomas-Fermi-Weizs\"acker equations. As a typical example, in our
calculations of the mixture with $N_B=3 \times 10^5$ and $N_F=1.5
\times 10^5$ atoms in an optical lattice of strength $V_0/h = 2.5
\,$kHz, from our converged solution we find that the rhs of
Eq.~(\ref{a4}) takes values $\sim 9$ ($\sim -80$) without (with) a
vortex line, while $U_H$ is $\sim 6 \times 10^6$ and $U_\mathrm{opt}$
is $\sim -5 \times 10^6$.

\end{document}